\documentclass{iau} 
\usepackage{graphicx}

\title[Class I methanol masers] 
{Class I methanol masers in low-mass star formation regions
}

\author[S.~Kalenskii~et~al.]   
{S.~Kalenskii$^1$, S.~Kurtz$^2$, P.~Hofner$^3$, P.~Bergman$^4$, C.M.~Walmsley$^5$,
 \and P.~Golysheva$^6$}

\affiliation{$^1$Lebedev Physical Institute, Astro Space Center, \\ 
84/32 Profsoyuznaya st., Moscow, GSP-7, 117997, Russia\\ email: {\tt kalensky@asc.rssi.ru}  \\[\affilskip]
$^2$Instituto de Radioastronomia y Astrofizika,  Universidad Nacional 
  Autonoma de Mexico, Morelia, Michoacan, Mexico \\email: {\tt s.kurtz@irya.unam.mx} \\[\affilskip]
$^3$National Radio Astronomy Observatory,
  1003 Lopezville Road, Socorro, NM 87801, USA \\ email: {\tt hofner\_p@yahoo.com} \\[\affilskip]
$^4$Onsala Space Observatory, Chalmers Univ. of Technology, 439 92 Onsala, Sweden \\ email: {\tt per.bergman@chalmers.se}\\[\affilskip]
$^5$ Dublin Institute for Advanced Studies, 31 Fitzwilliam Place, Dublin 2, Ireland (deceased)\\[\affilskip] 
$^6$119992, Universitetski pr., 13, Sternberg Astronomical Institute, Moscow University, 
Moscow, Russia\\ email: {\tt polina-golysheva@yandex.ru}
}

\pubyear{2017}
\volume{336}  
\jname{Astrophysical Masers: Unlocking the Mysteries of the Universe}
\editors{A. Tarchi, M.J. Reid \& P. Castangia, eds.}
\begin{document}

\maketitle

\begin{abstract}
We present a review of the properties of Class I methanol masers detected 
in low-mass star forming regions (LMSFRs). These masers, henceforth called LMMIs, 
are associated with postshock gas in the lobes of chemically active outflows 
in LMSFRs NGC1333, NGC2023, HH25, and L1157. LMMIs share the main properties
with powerful masers in regions of massive star formation and are a low-luminosity
edge of the total Class I maser population. However, the exploration of just these objects
may push forward the exploration of Class I masers, since many LMSFRs are located
only 200--300~pc from the Sun, making it possible to study associated objects 
in detail. EVLA observations with a $0.2''$ spatial resolution 
show that the maser images consist of unresolved or barely resolved spots with 
brightness temperatures up to $5\times 10^5$~K. The results are ''marginally'' 
consistent with the turbulent model of maser emission.

\keywords{ISM: jets and outflows, masers, radio lines: ISM.}
\end{abstract}

\firstsection 
\section{Introduction}

Bright and narrow maser lines of methanol (CH$_3$OH) have been found 
towards many star-forming regions. Methanol masers can be divided 
into two classes, I and II~(\cite[Menten, 1991b]{menten91b}), with each 
class characterized by a certain 
set of transitions. Class I maser transitions are the $7_0-6_1A^+$ 
transition at 44~GHz, $4_{-1}-3_0E$ transition at 36~GHz, $5_{-1}-4_0E$ 
transition at 84 GHz, $6_{-1}-5_0E$ at 132~GHz, $8_0-7_1A^+$ transition at 95~GHz etc., 
while Class II transitions are the $5_1-6_0A^+$ transition at 6.7~GHz, 
$2_0-3_{-1}E$ transition at 12~GHz, the series of $J_0-J_{-1}E$ 
transitions at 157~GHz, etc. The strongest Class I masers (usually called MMIs) emit at 44~GHz and demonstrate flux densities up to 800~Jy~(\cite[Haschick et al. 1990]{hmb}), while 
the strongest 
Class II masers (MMII) emit at 6.7~GHz and some of these achieve flux densities of $\sim 4000$~Jy~(\cite[Menten, 1991a]{menten91a}). Here we consider only MMIs; for a more
thorough description of their main properties see the contribution by Leurini \& Menten (this volume).

Until recently it was considered that methanol masers arise only in massive
star formation regions (MSFRs). But in the past few years several MMIs  
have been found in nearby low-mass star formation regions (LMSFRs) NGC1333, 
NGC2023, HH25, and L1157~(\cite[Kalenskii et al. 2006]{kalen06},
~\cite[Kalenskii et al. 2010a]{kalen10a},
~\cite[Kang et al. 2013]{kang13},
~\cite[Lyo et al. 2014]{lyo14}). The masers were 
detected in the Class I lines at 36~GHz, 44~GHz, 95~GHz, and 132~GHz.

\begin{figure}
\begin{center}
 \includegraphics[width=4.0in]{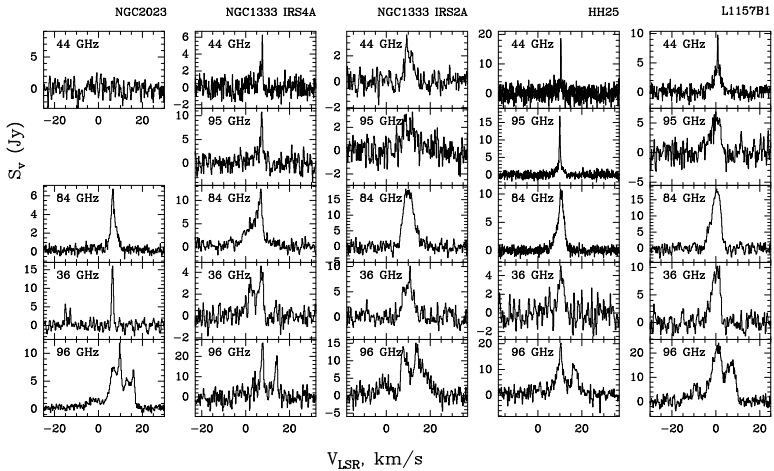} 
 \caption{LMMI spectra at 44, 95, 84, and 36~GHz, taken at the Onsala Space Observatory. 
Purely thermal $2_K-1_K$ methanol lines at 96~GHz are shown in the bottom row.}
   \label{fig1}
\end{center}
\end{figure}

\section{LMMI properties}
MMIs in LMSFRs (hereafter called LMMIs) have been studied using the 20-m Onsala radio telescope
and the KVN 21-m telescopes in a single-dish mode~(\cite[Kalenskii et al. 2010a]{kalen10a},
\cite[Kang et al. 2013]{kang13}, \cite[Lyo et al. 2014]{lyo14}).
In addition, four objects have been observed at 44~GHz with the VLA in 
the D configuration~(\cite[Kalenskii et al. 2010b]{kalen10b},
\cite[Kalenskii et al. 2013]{kalen13}) with a spatial resolution of $\sim 1.5''$.
The spectra of the lines observed at Onsala are shown in Fig.~\ref{fig1}. 
The main LMMI properties are discussed in \cite[Kalenskii et al.~(2013)]{kalen13}. They can be
summarised as follows:
\begin{itemize}
\item
All known LMMIs are associated with chemically active bipolar outflows,
where the gas-phase abundance of methanol is enhanced due 
to grain mantle evaporation.
VLA observations show that LMMIs are related to the shocked gas in the outflow lobes. 
\item
The known LMMIs are associated with clouds where the column densities of methanol are 
no less than $10^{14}$~cm$^{-2}$. \cite[Kalenskii et al. (2010a)]{kalen10a} 
suggested that MMIs can arise only when methanol column density 
is above this value.
\item
Flux densities of LMMIs do not exceed 18 Jy at 44~GHz and are lower
in the other Class I lines~(see~Fig~\ref{fig1}). However,  
LMMI luminosities at 44~GHz match the relation  between the protostar 
and maser luminosities $L_{\rm CH_3OH} = 1.71 \times 10^{-10} (L_{\rm bol})^{1.22}$, 
established for high- and intermediate-mass protostars by~\cite[Bae et al. (2011)]{bae}. 
\item
No variability at 44~GHz was detected in NGC1333I4A, HH25, or L1157 during
the time period 2004--2011.
\item
Radial velocities of most LMMIs  are close to the systemic velocities 
of associated regions. The only known exception is the maser detected at
36~GHz toward the blue lobe of the extra-high-velocity outflow 
in NGC 2023, whose radial velocity is 3.5~km~s$^{-1}$ lower than the systemic
velocity.
\end{itemize}
Thus, one can see that the main properties of LMMIs are similar to those of HMMIs. LMMIs 
are likely to be a low luminosity edge of the overall MMI population. 
Therefore the question arises, why should we 
study these few weak objects instead of focusing on much stronger MMIs in MSFRs? 
The answer is that the study of Class I methanol masers in LMSFRs  might be more 
straightforward compared to the study 
of the ''classical'' MMIs in MSFRs, because, in contrast to MSFRs,
LMSFRs are widespread and many of them are located only 200--300~pc from the Sun;
they are less heavily obscured in optical and IR wave ranges, and there are many 
isolated low-mass Young Stellar Objects (YSOs). We continue to study MMIs in LMSFRs
in order to better understand Class I methanol masers. Here we present the results 
of the observations of three maser sources performed at 44~GHz with the EVLA in 
the B configuration as well as CARMA observations of L1157 in the thermal lines 
of methanol $5_K-4_K$ at 241~GHz.

\section{New results}
\begin{figure}
 \includegraphics[width=3.1in]{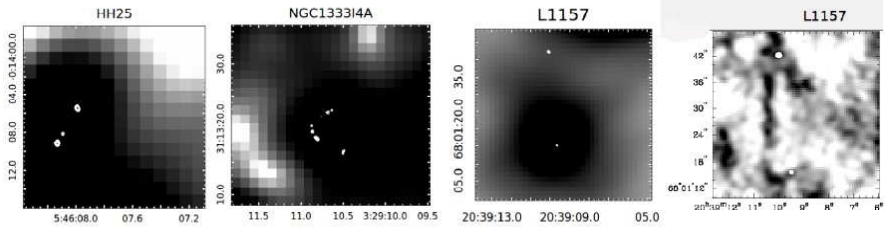} 
\vspace{-80mm}
 \caption{Three left panels: EVLA maps of methanol masers at 44~GHz (white dots) 
overlaid upon W1 
(HH25 and NGC1333I4A) and W4 (L1157) WISE images. Right panel: methanol masers overlaid upon 
the map of the $5_0-4_0A^+$ thermal emission in L1157.}
   \label{fig2}
\end{figure}

Spatial resolution of about $1.5''$, achievable at 44~GHz with the VLA in 
the D configuration, proved to be insufficient to resolve individual 
maser spots and measure their sizes and brightness temperatures. 
Therefore in 2013 we reobserved three LMMIs, HH25, NGC1333I4A, 
and L1157 at 44~GHz using EVLA in the B configuration, which provides 
an angular resolution of about $0.2''$ at 44~GHz. 
In addition, we observed L1157 in 
the $5_K-4_K$ thermal lines of methanol at 241~GHz with the antenna 
array CARMA in the C configuration with an angular resolution of $\approx 1''$.

\begin{figure}
 \includegraphics[width=5.2in]{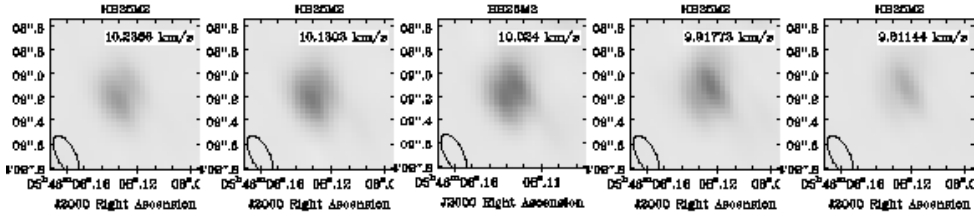} 
\vspace{-157mm}
 \caption{Example of a double spot: channel map of HH25M2 at 44~GHz.}
   \label{fig3}
\end{figure}

A collection of the overall maps of the three observed sources at 44~GHz 
is shown in Fig.~\ref{fig2}. 
The maps show that each source consists of several spots. Hereafter M1 means 
the strongest spot in the region, M2 is the second strongest spot etc.
Deconvolved spot sizes vary from $\sim$0.10$''$--0.15$''$ for the stronger spots to $\sim$0.10$''$--0.3$''$ for the weaker spots (30--45~AU and 30-90~AU, respectively).

The brightness temperatures of the strongest spots are as high as $5\times 10^5$~K.

Maser spots in NGC1333I4A form an arc around a NIR object clearly seen in the W1 and W2 WISE maps.

The maps of individual
spots show that many of them can be decomposed into two unresolved 
compact subspots. In these cases the brightest subspot 
is denoted subspot a, the second brightest subspot, subspot b.
An example of such double spot is shown in
Fig.~\ref{fig3}, which exhibits the channel map of the second brightest 
spot in HH25 (HH25M2).
Among the spots that demonstrate double structures are L1157M1 and M2, 
NGC1333I4AM2 etc. 

An interesting result is the detection of unresolved spots demonstrating broad
($\gtrsim 3-5$~km~s$^{-1}$) spectral lines. Their fluxes are about 0.1--0.2~Jy,
which corresponds to brightness temperatures $\gtrsim 1000$~K. Thus, in spite
of large linewidths, these objects are probably weak masers. 

\section{Discussion}
According to the most popular maser model, compact maser spots arise in 
extended turbulent clumps because in a turbulent velocity field the coherence 
lengths $l$ along some lines are increased. If masers are associated with 
turbulence, the map appearance depends on the maser 
regime. Saturated regime of maser amplification is characterized by a large number of spots 
of comparable intensity, while the unsaturated maser amplification results
in a small number of bright spots~(\cite[Strelnitski et al. 2017]{strelnitski}). 
The map of maser emission in L1157 with only two maser spots (Fig.~2) favors 
the unsaturated regime of maser amplification, which was studied 
by~\cite[Sobolev et~al.~(1998)]{sob98}. 

One of the main parameters of the model by Sobolev et al. is $\tau_0$, the absolute value 
of optical depth at the center of the inverted line when there is
no turbulence in the cloud. From the intensities of thermal lines $5_K-4_KA^+$ toward 
M1, observed with CARMA, we estimated N$_{\rm CH_3OH}$ ($\sim 10^{16}$~cm$^{-2}$) 
and $\tau_0$ at~44~GHz ($\sim 12$). From Table~1 
of~\cite[Sobolev et al. (1998)]{sam98} we estimated that the optical depth at 44~GHz 
toward M1 is $\boldmath \tau^{44} \sim 7-8$ and $T_{br}$ at this frequency 
$\sim 10^4$~K, much lower than the observed one. However, an increase of N$_{\rm CH_3OH}$ 
by a factor of less than 2 makes it possible
to achieve the observed brightness temperature.
Thus, the turbulence model is in ''marginal'' agreement with the observations. 

SVK acknowledges the support of the Russian Foundation for Basic Research (project no. 15-02-07676).



\end{document}